\documentclass[conference]{IEEEtran}
\IEEEoverridecommandlockouts
\usepackage{cite}
\usepackage{amsmath,amssymb,amsfonts}
\usepackage{algorithmic}
\usepackage{graphicx}
\usepackage{textcomp}
\usepackage{xcolor}
\usepackage{todonotes}
\def\BibTeX{{\rm B\kern-.05em{\sc i\kern-.025em b}\kern-.08em
    T\kern-.1667em\lower.7ex\hbox{E}\kern-.125emX}}
    \makeatletter
\def\ps@IEEEtitlepagestyle{%
\def\@oddfoot{\mycopyrightnotice}%
\def\@evenfoot{}%
}
\def\mycopyrightnotice{%
{\footnotesize 978-1-6654-3288-7/21/\$31.00~\copyright~2021 IEEE\hfill}
\gdef\mycopyrightnotice{}
}
\begin{document}

\title{Visually Exploring Multi-Purpose Audio Data\\
\thanks{This work is supported by the National Science Foundation, award 2100874.}
}

\author{\IEEEauthorblockN{David Heise}
\IEEEauthorblockA{\textit{Lincoln University} \\
\textit{Jefferson City, MO, USA} \\
heised@lincolnu.edu
\and
\IEEEauthorblockN{Helen L. Bear}
\IEEEauthorblockA{\textit{Queen Mary University of London} \\
\textit{London, UK}\\
h.bear@qmul.ac.uk}
}
}

\maketitle

\begin{abstract}
We analyse multi-purpose audio using tools to visualise similarities within the data that may be observed via unsupervised methods.  The success of machine learning classifiers is affected by the information contained within system inputs, so we investigate whether latent patterns within the data may explain performance limitations of such classifiers.  We use the visual assessment of cluster tendency (VAT) technique on a well-known data set to observe how the samples naturally cluster, and we make comparisons to the labels used for audio geotagging and acoustic scene classification.  We demonstrate that VAT helps to explain and corroborate confusions observed in prior work to classify this audio, yielding greater insight into the performance -- and limitations -- of supervised classification systems.  While this exploratory analysis is conducted on data for which we know the ``ground truth" labels, this method of visualising the natural groupings as dictated by the data leads to important questions about unlabelled data that can help the evaluation and realistic expectations of future (including self-supervised) classification systems.
\end{abstract}


\section{Introduction}
Acoustic scene classification (ASC) \cite{Chen2019} is an established research task, classifying audio recordings according to a list of scene labels.  Audio geotagging is a related problem where the labels correspond to the geographical location where the recording was taken \cite{bear2019city}. These are fundamentally different machine learning problems, but they can be investigated with the same data. To date, audio geotagging has proven more difficult, suggesting that a data set may contain information that may be more or less salient for a given task.

In multi-purpose data, is the discriminating information equally easy to learn for different tasks? For an ASC model which is trained on data all recorded in a single city, its ability to discriminate scenes is not confounded by location variation. (Other confounding variables \textit{are} present; this is just an example.) Yet, in order for an ASC model to be useful in the real-world, it has to predict accurately for any location. 
More complex data is needed to train robust systems, and this comes with the cost of confounding variables. 

This work focuses on understanding --~and seeks to explain~-- the information in the data before machine learning is undertaken.  State-of-the-art and conventional methods for these problems use a traditional pipeline:  1)~data collection, 2)~feature extraction, 3)~model training, 4)~classification/prediction, and 5)~evaluation. Whilst the majority of recent research focuses on an end-to-end training (deep learning) \cite{Suh2020}, this precludes understanding or interpretation at mid-way points, which could in turn make model training more difficult than necessary. 

We are motivated by the inherent assumption that data from the same class are discernible from data in other classes in some way, however complex that function might be. Class labels are often from a taxonomy that does not consider the data; for example, in the data set used here, \textit{airport} is so labeled not because of the distinctive sounds in that environment, but rather because of the place and type of activity that happens there. The goal of classifiers is to predict these labels irrespective of the complexity of the data (by learning hidden patterns within it), but there is value in identifying the nature of that complexity.  In multi-purpose data, tasks have different inter-class and intra-class similarities, and this has a direct impact on the difficulty of matching data inputs to annotations (and thus discrimination between classes). 


We are further motivated by Detection and Classification of Acoustic Scenes and Events (DCASE) challenge reports (surveyed in \cite{mesaros2019acoustic}). There is a pattern of participants training audio machine learning (ML) models and spending significant time optimising model parameters for minute performance gains. In many papers, there is no discussion of why or how the algorithms fail for certain predictions.  This is unsurprising, given that most participants use ``black-box" machine learning approaches which are inherently unexplainable.  Also, in a lot of prior work, predictions produce misclassifications between class labels; we wonder if this may be due to similarities in the raw data (between recordings that have different annotations).  This leads to the questions:
\begin{itemize}
\item does the data contain sufficient discriminatory data to separate into the different sets of labels for different tasks, despite the confounding factors in the data?, and
\item are certain pairs or sub groups of classes getting confused, e.g. \textit{metro} and \textit{metro\_station}?
\end{itemize}




\section{Background}
\label{sec:background}
\subsection{Prior work}
In environmental audio processing research there are many tasks: ASC \cite{zheng2019acoustic}, sound event detection (SED) \cite{tak2017novel}, sound event localisation \cite{adavanne2018sound}, audio geotagging \cite{kumar2017audio}, audio tagging \cite{fonseca2018general}, and audio captioning \cite{ccakir2020multi} to list but a few. All of these, subject to labelling overhead, could be undertaken with the same dataset -- that is, a single collection of recorded sound scenes. 
At present, many sound scene data sets exist but are only annotated for single tasks, such as: URBAN-SED \cite{salamon2017scaper}, Audioset \cite{gemmeke2017audio}, and the various DCASE challenge datasets \cite{dekkers2018dcase,adavanne2019multi,stowell2015detection}. With the popularity of deep learning approaches dominating state-of-the-art solutions for sound scene analysis, saliency maps can be used for determining which parts of the sound scenes are the most useful for a neural network to predict classes \cite{simonyan2013deep} -- but this does not indicate whether the training data contains enough information to discriminate between classes.  The training data sets used for the state-of-the-art sound scene analysis are currently real-world recordings. These recordings are influenced by many factors in addition to the location or scene (e.g., time of day or day of the week). 

The data explored here is used for two tasks in the literature: ASC and audio geotagging. In the DCASE 2020 ASC challenge, the participant with the highest accuracy \cite{Suh2020} used data augmentations with a triple parallel ResNet model. There is no verification of the effect of the augmentations on the inputs before model training. Prior work in audio geotagging is limited. In \cite{bear2019city}, the highest accuracy is achieved with a multi-task neural network which predicts both cities and scenes.

\subsection{VAT analysis}
\label{sec:vact}
We use Bezdek and Hathaway's visual assessment of (cluster) tendency (VAT) \cite{bezdek2002vat}. VAT is a simple unsupervised method to visualise salient clusters of data points within a dataset, and many variants and extensions have been developed \cite{Kumar2020}. A key benefit of this method over others is that it does not require an estimate $k$ value as a target number of clusters. This visualisation can help researchers to see if feature extraction methods are selecting useful features or discarding them, the latter making training an accurate classifier more difficult. To date, VAT has been rarely applied to audio data, and only on speech utterances \cite{RajendraPrasad2017}. 

VAT calculates the distance between all pairs of data points.  (We used Euclidean distance, though VAT is not constrained to this choice.)  The two furthest data points are identified, allowing one to be selected as first in the reordering of points.  Each subsequent point in the reordering is selected by choosing the point that is nearest to any point that has already been reordered, and the resulting sequence of points is the same as the order that points would be added to a minimal spanning tree of the complete graph per Prim's algorithm \cite{Havens2009}.  The intensity values of the resulting ordered dissimilarity matrix are the distance between all pairs of recordings.  Rendered as a grayscale image, the number of dark boxes visible on the diagonal \textit{may} suggest the number of clusters in the data. This suggestion of how many distinct classes are in the data is based solely on the data itself with no bias for any labelling system.

SpecVAT is an extension to VAT based on spectral graph theory \cite{WangSpecVAT2008}.  This extension attempts to better deal with complex cluster structure by first mapping the dissimilarity matrix $D$ from VAT into a weighted affinity matrix (weighted according to local neighborhood).  This is followed by spectral decomposition of the normalized Laplacian of the weighted affinity matrix, resulting in an embedded feature space $D'$ based on the $k$ largest eigenvectors.  The embedded space $D'$ is then reordered using VAT.  The optimal value for $k$ may be deduced by A-SpecVAT \cite{WangSpecVAT2010} or other means, seeking a value that yields the clearest dark boxes on the diagonal.  Images produced by SpecVAT tend to yield clearer structure within complex data than those produced directly by VAT.

There are two important decisions here: the choice of distance metric, and the choice of feature space.  Euclidean distance appeared to perform more robustly than other measures attempted, and the log mel spectrogram feature is a commonly used for both audio tasks. 
\section{Method}
\label{sec:method}
In this work the DCASE 2018 ASC Task 1a \cite{Mesaros2018_DCASE} dataset (described in Sec~\ref{ssec:data}) is used. Beginning with the data divided into subsets of scene specific or city specific subsets, VAT analyses (described in in Sec~\ref{sec:vact}) are undertaken directly on the log mel spectrogram features, producing $16$ ordered dissimilarity images for analysis one. From these images, both the salient groups of recordings are counted and more easily confused labels are identified. Also plotted in the visualisations is the shortest path as a stacked bar chart to make inter-class comparisons by colour-coding the desired labels. In these stacked bars, larger blocks of the same colour show that recordings within the same class are more similar in data space to each other than they are similar to recordings of other classes.

\begin{figure*}[htbp]
\centerline{\includegraphics[width=1.5\columnwidth]{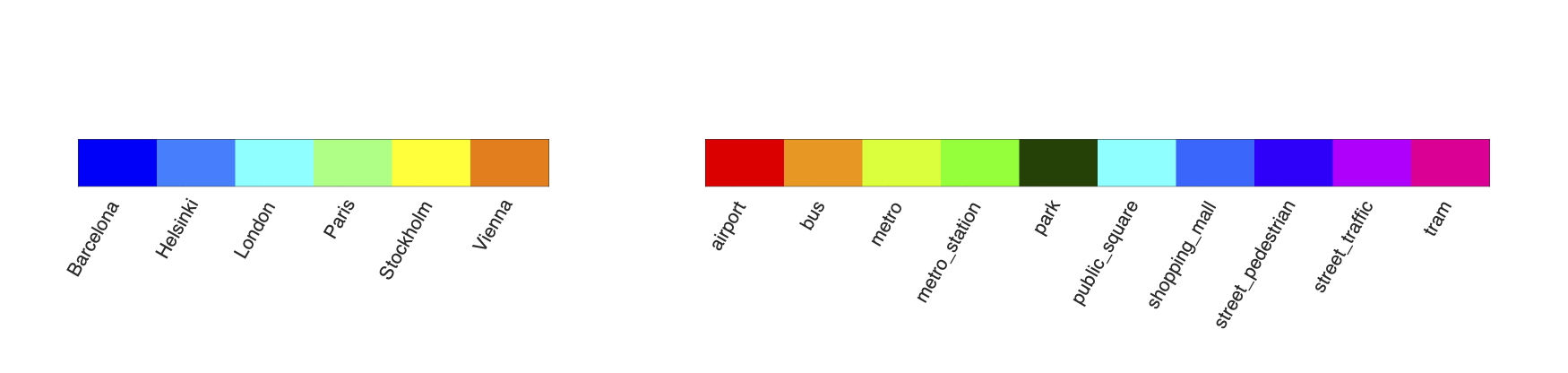}}
  \centering
  \includegraphics[width=2.1\columnwidth]{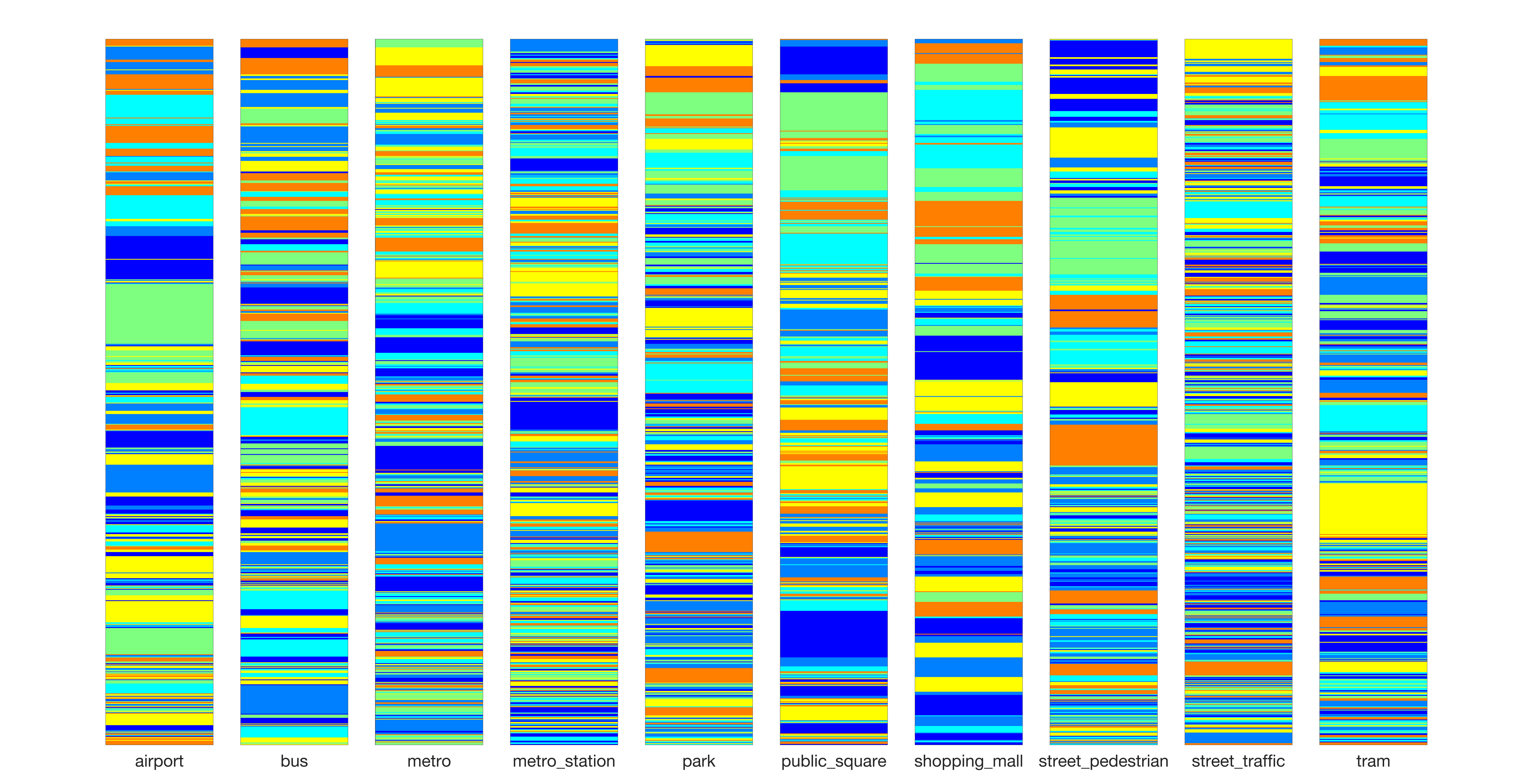}
  \centerline{(a) Label blocks of cities in single scene data }
  \includegraphics[width=1.3\columnwidth]{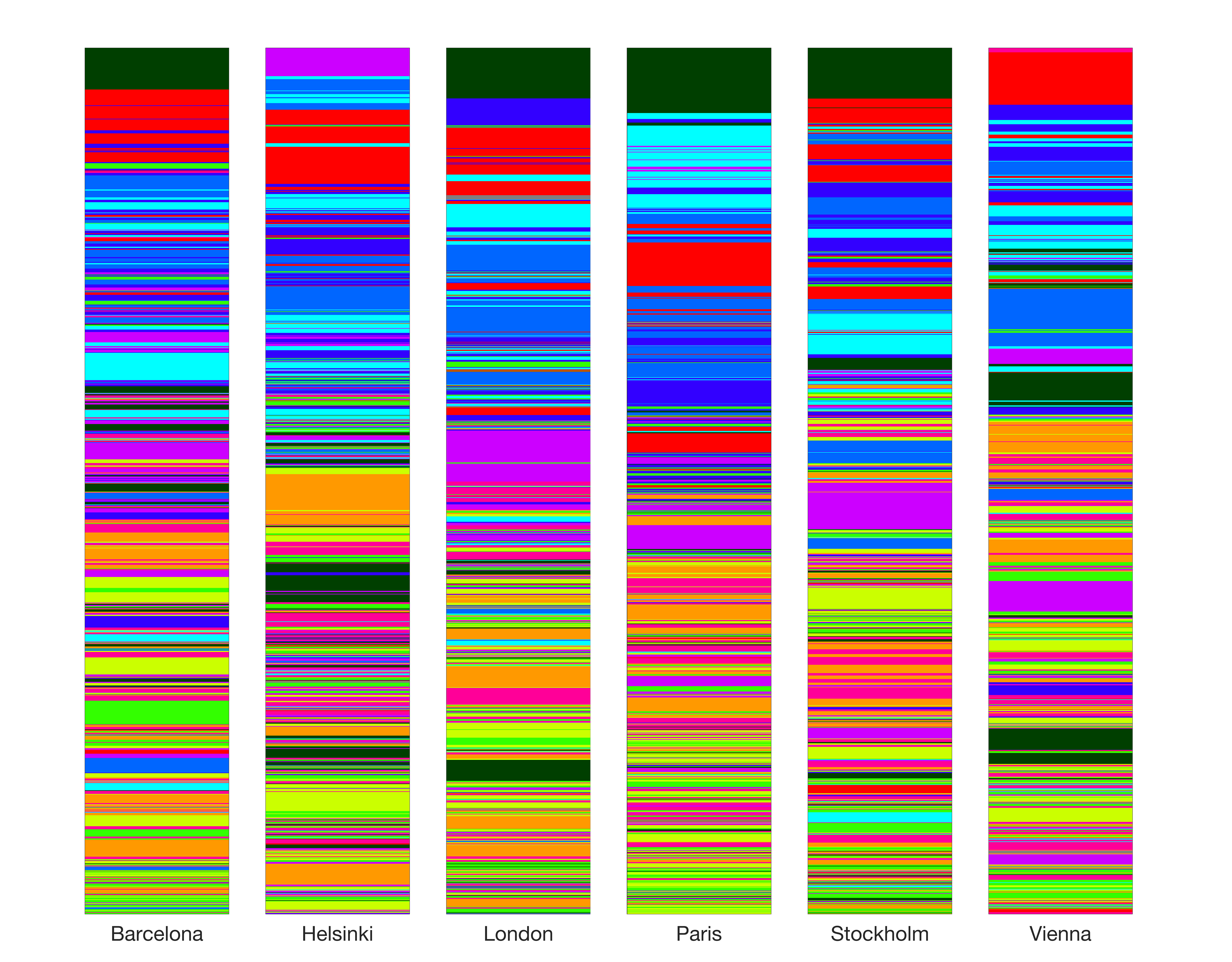}
  \centerline{(b) Label blocks of scenes in single city data}
\caption{A VAT analysis for each subset of data. Each column represents a subset, e.g., only airport recordings for all six cities, or only Stockholm recordings for all ten scenes. The labels are ordered by the rows of the VAT analysis (unsupervised) and colour coded to the annotations. Chunks of the same colour show where recordings having the same label are most similar to each other. A particular colour distributed throughout a stack indicates that the intra-class variation of that class is significant and will be difficult to discriminate from other annotations.}
\label{fig:stacked}
\end{figure*}

The second analysis uses all the data; that is, with the confounding variables (all cities and all scenes).  Using the cluster count extraction (CCE) method of Sledge et al. \cite{Sledge2008}, this method works by first thresholding the VAT image using Otsu's algorithm and then constructing a histogram of an off-diagonal slice of the image.  For this work, default CCE parameters are used except for the value $b$, which is set to one half of the maximum value of the computed histogram (rather than $b = 0$).  The automatic counts yielded by the algorithm are given in Table~\ref{tab:clustersByClass}. 

\subsection{Data}
\label{ssec:data}
Audio data is from the DCASE 2018 ASC subtask 1A \cite{Mesaros2018_DCASE}. This data was recorded from ten different scenes (\textit{airport}, \textit{bus}, \textit{metro}, \textit{metro\_station}, \textit{park}, \textit{public\_square}, \textit{shopping\_mall}, \textit{street\_pedestrian}, \textit{street\_traffic}, and \textit{tram}) in six cities (\textit{Barcelona}, \textit{Helsinki}, \textit{London}, \textit{Paris}, \textit{Stockholm}, and \textit{Vienna}). The data was labelled for ten sound scenes and relabelled for cities in \cite{bear2019city}. It was recorded between 9am and 9pm on different days of the week in varied weather. 

Each acoustic scene has 864 ten-second segments (giving $8640$ segments across ten scenes). These were recorded using a binaural Soundman OKM II Klassik/studio A3 electret in-ear microphone and a Zoom F8 audio recorder using $24$-bit resolution.
Using librosa \cite{mcfee2015librosa}, log mel spectrogram features are extracted using $128$ mel bands and a $2048$-point STFT. The input sampling rate was $22050$ samples per second and the hop length was $512$. Each recording was transformed from a $431$ column matrix into a single vector by taking the feature-wise mean over all time frames. 

\section{Analysis One:  By Class}
\label{sec:analysis1}
Fig.~\ref{fig:stacked} is a series of stacked bars which represent the annotated labels for each recording, reordered by VAT. Each stack in Fig.~\ref{fig:stacked}a represents one of the ten scenes; each stack in Fig.~\ref{fig:stacked}b represents a single city (as per audio geotagging labels). For each set, recordings are represented by one row in the stack, neighboured by its most similar recordings. Blocks of colour show similarity within a class because recordings from the same class are neighbours on the path. 

In Fig.~\ref{fig:stacked}a, city blocks are visible within some single scenes. 
In \textit{airport} and \textit{shopping\_mall}, one can see consistent chunks of colour (with relatively few short chunks), indicating groupings of recordings which map to the city annotations.  There is something distinctive about the \textit{street\_pedestrian} recordings taken in Paris, as they also group together.  
One interesting point is from \textit{airport}, that most of the Paris recordings are grouped together. This may be explained by the fact that Charles de Gaulle Airport has a unique announcement tone which can be heard in a number of the recordings. 
In the \textit{street\_traffic} stack, there are very few chunks of colour of any noticeable size, and the colours are distributed throughout the stack. This shows that street traffic sounds are very similar regardless of the city.

In Fig.~\ref{fig:stacked}b, \textit{public\_square}, \textit{shopping\_mall}, \textit{airport}, and \textit{street\_pedestrian} cluster together for all cities, as seen by the combination of reds and blues near the top of each stack. \textit{metro}, \textit{metro\_station} and \textit{bus} are also often in close proximity; this might be due to similar activities in these places.  Overall, \textit{park} and \textit{airport} are the most distinctive scenes when VAT is run on a single city.

\begin{figure*}[!htb]
\begin{minipage}[b]{0.5\linewidth}
  \centering
  \centerline{\includegraphics[width=\columnwidth]{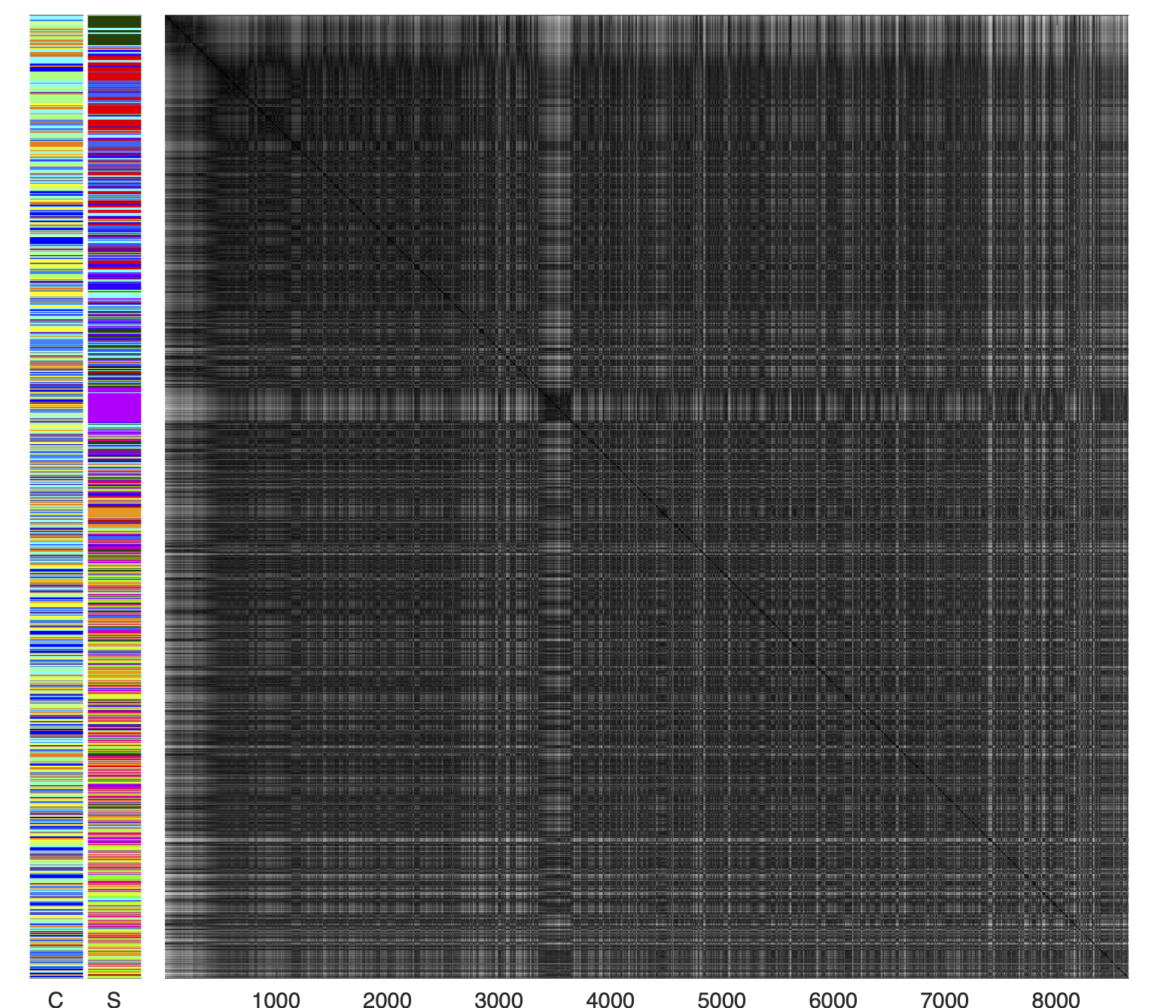}}
\end{minipage}
\begin{minipage}[b]{0.5\linewidth}
  \centering
  \centerline{\includegraphics[width=\columnwidth]{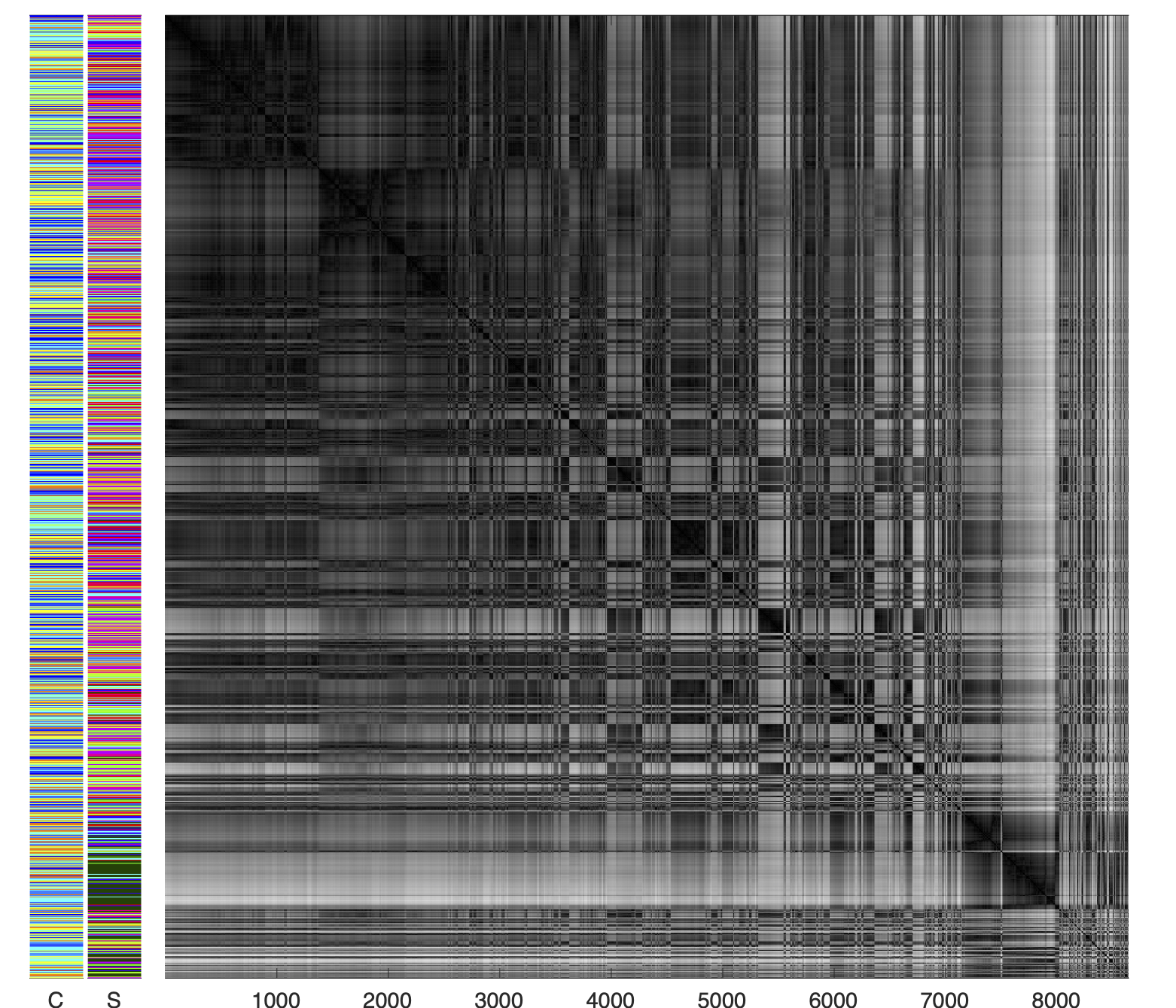}}
\end{minipage}
\\
\\
\begin{minipage}[b]{0.5\linewidth}
  \centering
  \centerline{\includegraphics[width=0.75\columnwidth]{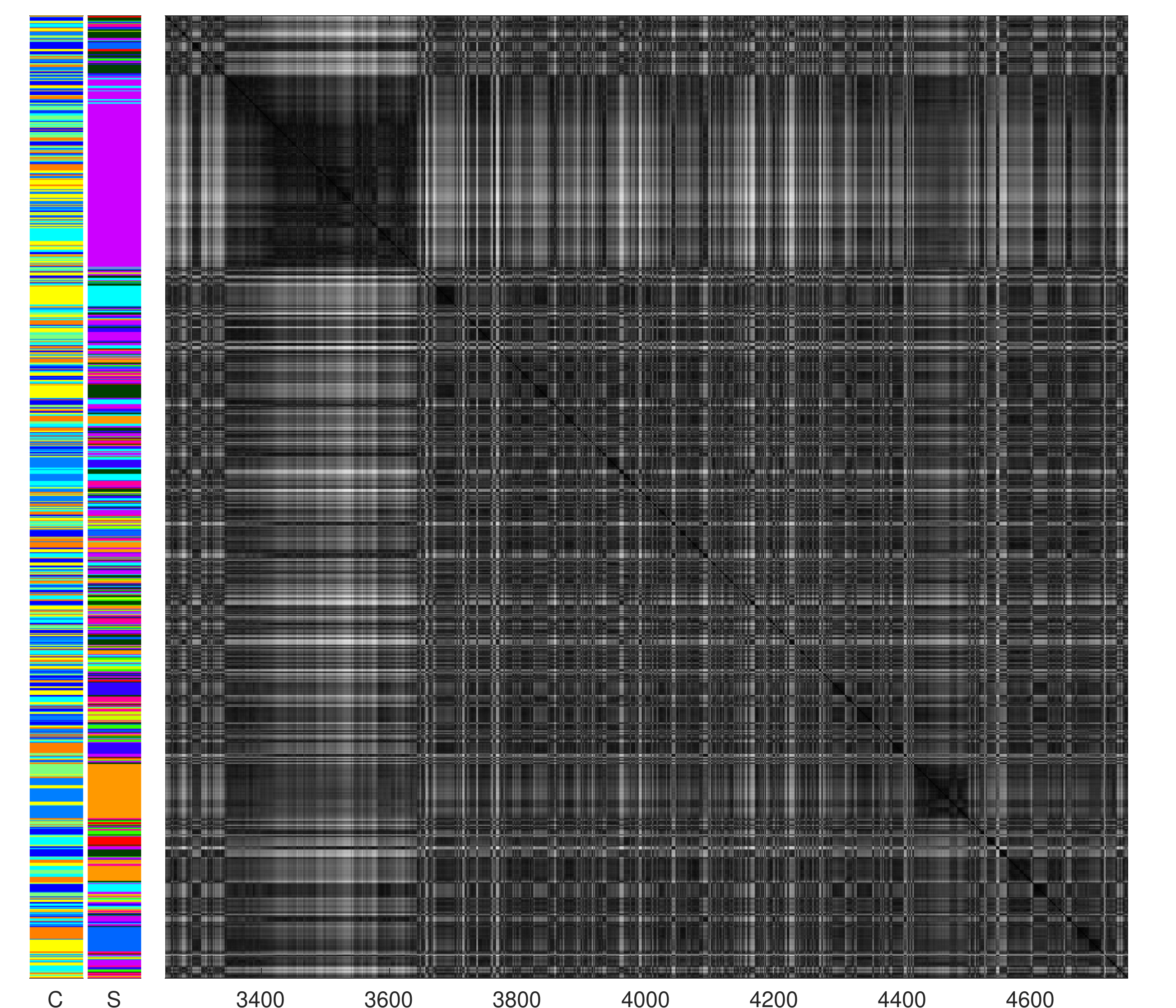}}
\end{minipage}
\begin{minipage}[b]{0.5\linewidth}
  \centering
  \centerline{\includegraphics[width=0.75\columnwidth]{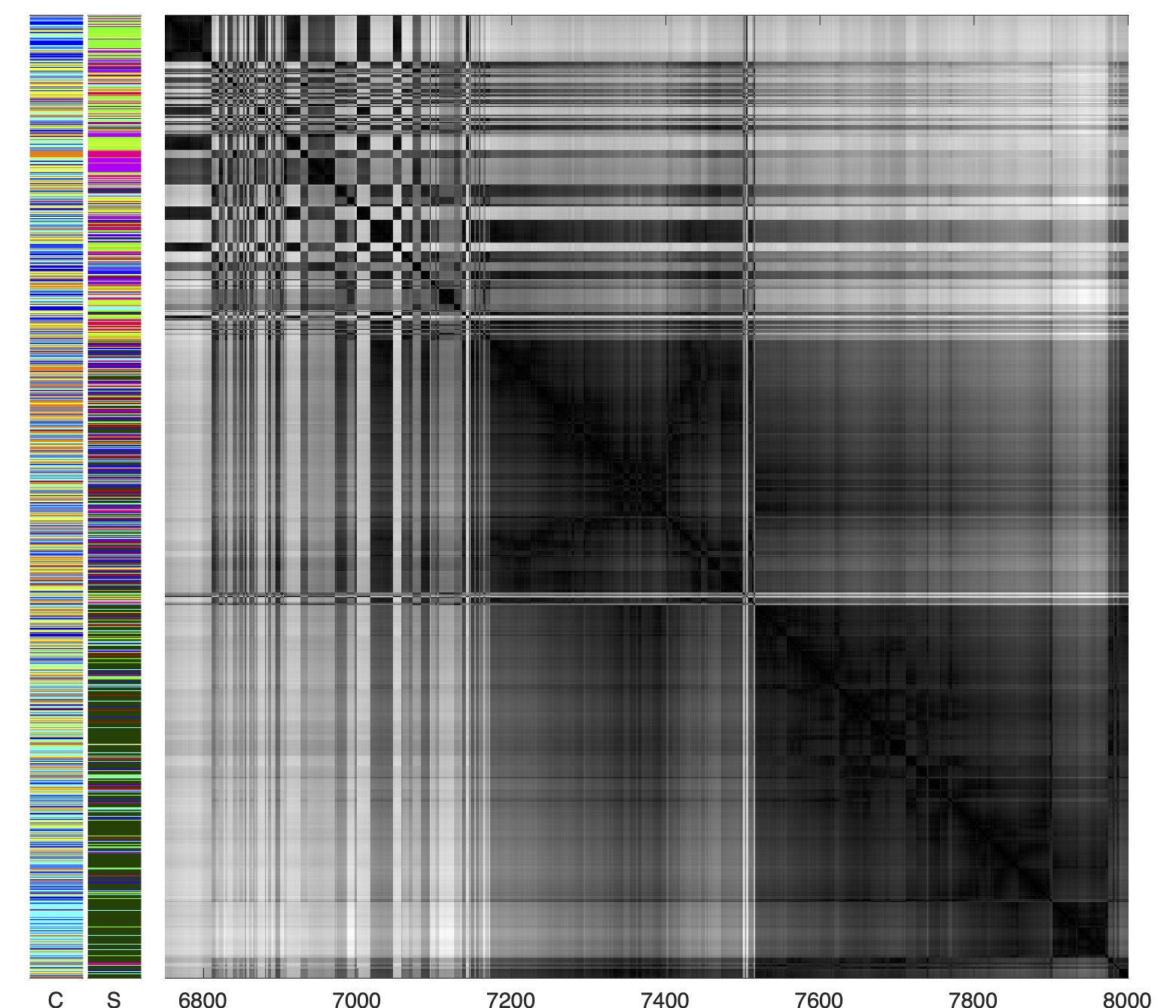}}
\end{minipage}

\caption{A series of ordered dissimilarity matrices produced by VAT and SpecVAT on multi-purpose audio data. At left of each dissimilarity image are stacked bars showing labels for  city labels (`C') and scene labels (`S'). Clusters of very similar recordings are darkened squares on the diagonal. \\
\textbf{Top left:} a VAT ordered dissimilarity image created from all data in \cite{Mesaros2018_DCASE}.  There are $42$ clusters on the diagonal counted by CCE.\\ 
\textbf{Bottom left:}  a zoomed-in section for recordings 3251-4750.\\
\textbf{Top right:} a SpecVAT (number of eigenvectors $k=3$) ordered dissimilarity image created from all data in \cite{Mesaros2018_DCASE}.\\
\textbf{Bottom right:} a zoom of indices 6751-8000 from the SpecVAT image above.  \\ Apparent clusters seem to be comprised of data points with majority same label, though these chunks of color are interspersed with data points having other labels -- potentially explaining some sources of misclassification from ML models.}
\label{fig:vats}
\end{figure*}

Table~\ref{tab:clustersByClass} shows the number of estimated clusters in VAT images for each of the stacked bars in Fig~\ref{fig:stacked}. These cluster counts are estimates derived using the CCE algorithm mentioned in Sec.~\ref{sec:method}. When data is grouped by city, there are approximately ten clusters.  This is inspiring given the target ten scene labels, but it would be na\"{\i}ve to assume that the suggested partitions of the data map exactly to those target labels.  It is important to recognize that this observation excludes the confounding factor of recordings from different cities.  When data is grouped by scene, there is greater variability in the number of clusters. This suggests that discriminating city recordings, even within a single type of scene, is a more complex task where the type of scene will have an effect on how to discriminate cities. 

\begin{table}[!htb]
\centering
    \caption{Estimated cluster count by city and scene}
    \begin{tabular}{l|c||l|c}
    City & \# & Scene & \# \\ 
    \hline
    Barcelona & 10 & airport & 8\\
    Helsinki & 9 & bus & 2\\
    London & 11 & metro & 3\\
    Paris & 14 & metro\_station & 4\\
    Stockholm & 14& park & 12\\
    Vienna & 11 & public\_square & 14\\
    & & shopping\_mall & 15\\
    & & street\_pedestrian & 11\\
    & & street\_traffic & 5\\
    & & tram & 13\\
    \end{tabular}
    \label{tab:clustersByClass}
\end{table}

\section{Analysis Two:  All Data}
\label{sec:analysis2}
Analysis two uses the whole dataset.  Fig.~\ref{fig:vats} contains a series of VAT ordered dissimilarity matrices, where dark boxes on the diagonal suggest possible clusters.  To the left side of each matrix are two ordered stacked bars for cities (`C') and scenes (`S'), labelled with the same ground truth colours as in Fig.~\ref{fig:stacked}.
CCE counts $42$ clusters in Fig.~\ref{fig:vats} (top left). This is over four times the possible target annotated labels (six or ten). While these $42$ clusters are not clear in the full VAT, looking at the zoomed-in portion (bottom left) there are clear dark blocks that correspond to particular scenes (e.g., \textit{street\_traffic}, \textit{public\_square}, and \textit{bus}). This is consistent with Fig.~\ref{fig:stacked}, suggesting useful clusters for these tasks in the presence of confounding variables.  Therefore, the problem is not as simple as a 60-class classification; rather, it may be a multi-label challenge. This suggestion is supported by results in \cite{bear2019city} where results showed that multi-task learning for jointly classifying scenes and cities achieved the greatest accuracy. 

\begin{figure*}[htb]
\begin{minipage}[b]{0.5\linewidth}
  \centering
  \centerline{\includegraphics[width=\columnwidth]{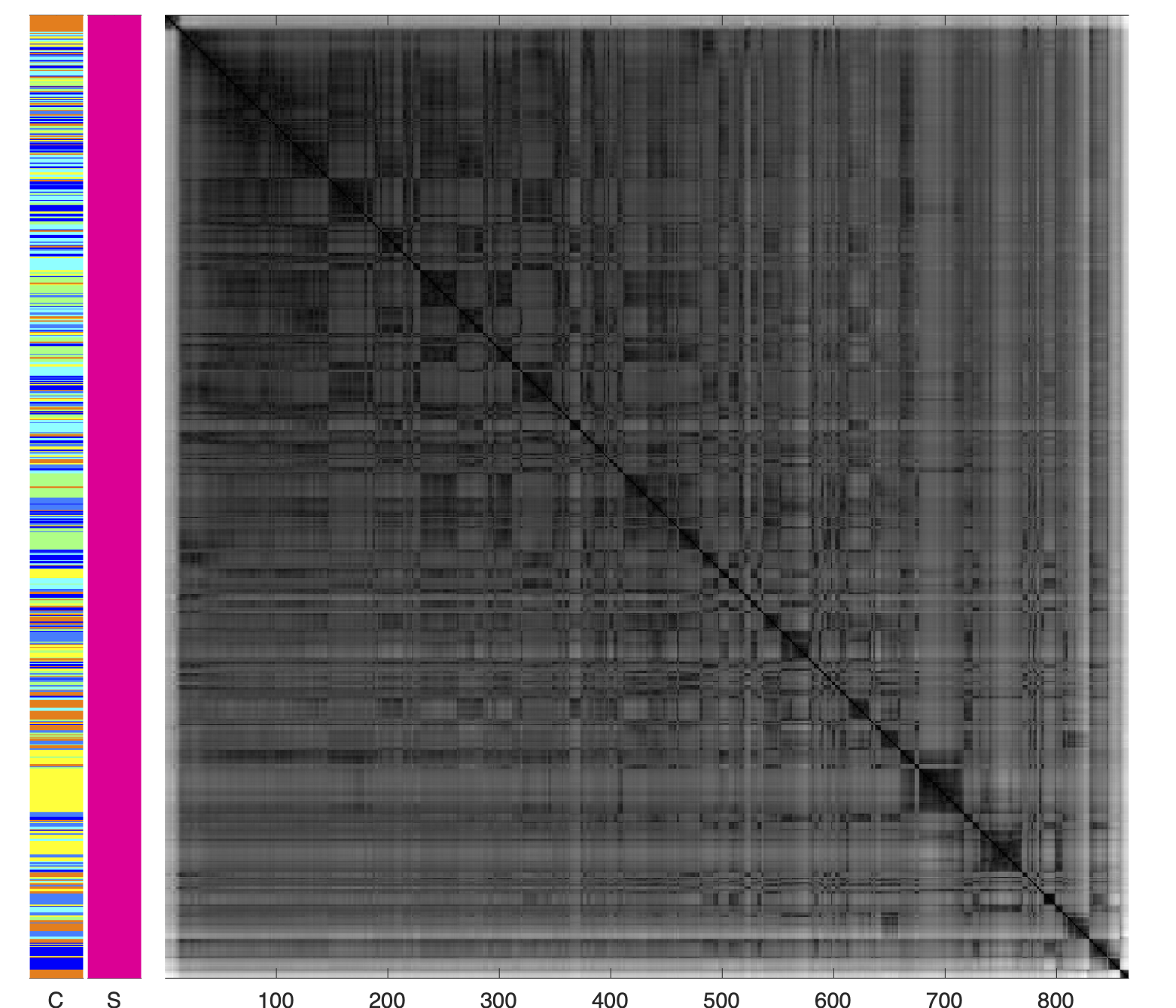}}
\end{minipage}
\begin{minipage}[b]{0.5\linewidth}
  \centering
  \centerline{\includegraphics[width=\columnwidth]{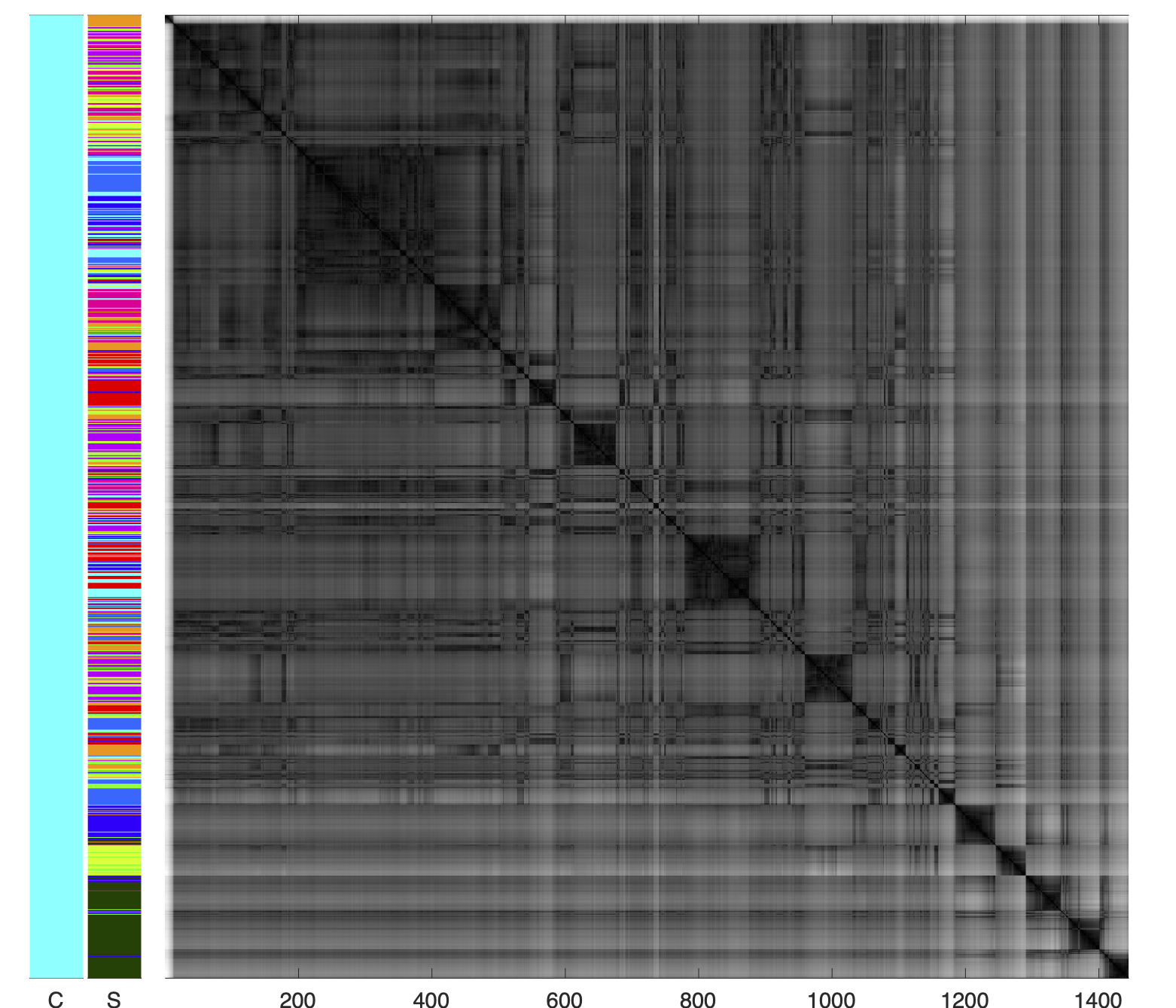}}
\end{minipage}
\caption{Single task SpecVAT for \textit{tram} recordings (left) and \textit{London} recordings (right).}
\label{fig:vats3}
\end{figure*}


Fig.~\ref{fig:vats} (top right) shows the same data run through SpecVAT (eigenvectors $k=3$ as suggested by A-SpecVAT \cite{WangSpecVAT2010}). Again, the city and scene labels are on the left and a zoom is on the bottom right. The dark blocks on the diagonal are more obvious to see, and there appears to be less noise off the diagonal line. Whilst there are fewer contiguous label chunks for both scenes and cities, if each dark block is considered a local `area' within the scene stack, then each area contains a prevalence of one colour, even if the order is interrupted by a few other colours. Based on these observations, the final analysis uses SpecVAT. 

\section{Analysis Three:  Separate Audio Tasks}
This final analysis demonstrates the difference when known confounding factors are removed -- that is, when multi-purpose data is simplified for a single task. To do so, Fig.~\ref{fig:vats3} presents two final SpecVAT dissimilarity matrices. On the left is the SpecVAT image using only \textit{tram} recordings as could be used for city classification (audio geotagging), and on the right is the SpecVAT image using only \textit{London} recordings that could be used for ASC.  

The key message from these figures is that the dark squares on the diagonals are distinct and there are contiguous blocks of colour on the label stacks.  These squares/blocks are detected in a purely unsupervised manner.  The as-yet unsolved problem is how one gets from the unsupervised blocks to corresponding annotations, particularly where the neighboring recordings do not have matching labels. This is future work.


\section{Conclusions}
\label{sec:recs}
In this work, multi-purpose audio data have been analysed to address the question of whether the same data contains enough information to discriminate into different labelling schemes. Whilst this work focused on ASC and geotagging, other tasks like sound event detection could also benefit from this approach. Using VAT has revealed that there is some structure which relates to the different target classifications. This structure is further visible with SpecVAT which appears to denoise standard VAT. With our data, getting to scene labels appears an easier task than city labels. 

In practical terms, this structure can also explain some common misclassifications seen in prior predictive work. For example, in \cite{bear2019city} the classes commonly confused by the classification models were the same as detailed in Sec.~\ref{sec:analysis1}, such as \textit{airport} and \textit{shopping\_mall}. Further VAT analysis reveals class pairings that are similar in the data space but not assumed to be similar by their labels (such as the \textit{airport} / \textit{public\_square} / \textit{shopping\_mall} / \textit{street\_pedestrian} confusions noted in Sec.~\ref{sec:analysis1}).

This work enables researchers to better understand their input data. Using visualisation techniques such as VAT and SpecVAT to understand the relations within the data structure, and comparing the results to target labels in the research problem, will help post-classification performance evaluation. 
Our data-driven approach to exploring patterns within the data set is akin to the ideas behind self-supervised classification methods. These models learn well without training labels, suggesting that the raw data contains sufficient discriminatory information. These models, however, do not yet show us the data structure the classifier is learning without significant training effort and the utilisation of further explainability models. Our approach is much simpler and can inspire a model design \emph{before} it is trained. 



\bibliographystyle{IEEEbib}
\bibliography{refs}

\end{document}